\documentclass[
reprint,
superscriptaddress,
nofootinbib,
 amsmath,amssymb,
]{revtex4-1}

\pdfoutput=1
\usepackage{float}
\usepackage{xcolor}
\usepackage{graphicx}
\usepackage{dcolumn}
\usepackage{bm}
\usepackage[colorlinks=true, linkcolor=blue, citecolor=blue, urlcolor=blue]{hyperref}
\newcommand{\ZIB}{Zuse Institute Berlin, Takustra{\ss}e 7, 14195 Berlin, Germany}
\newcommand{\JCM}{JCMwave GmbH, Bolivarallee 22, 14050 Berlin, Germany}

\usepackage{graphicx}
\usepackage{booktabs}
\usepackage{tabularx}

\newcommand{\mytoprule}{\specialrule{0.1em}{0.2em}{0.2em}}
\newcommand{\mymidrule}{\specialrule{0.1em}{0.2em}{0.2em}}
\newcommand{\mybottomrule}{\specialrule{0.1em}{0.2em}{0.2em}}

\begin{document}

\title{Quasinormal mode expansion of optical far-field quantities}

\author{Felix Binkowski}
\affiliation{\ZIB}
\author{Fridtjof Betz}
\affiliation{\ZIB}
\author{R\'emi Colom}
\affiliation{\ZIB}
\author{Martin Hammerschmidt}
\affiliation{\JCM}
\author{Lin Zschiedrich}
\affiliation{\JCM}
\author{Sven Burger}
\affiliation{\ZIB}
\affiliation{\JCM}

\begin{abstract}
\vspace{0.1cm}
Quasinormal mode (QNM) expansion is a popular tool to analyze light-matter
interaction in nanoresonators. However, 
expanding far-field quantities such as the energy flux  is an open problem because
QNMs diverge with an increasing distance to the resonant systems.
We introduce a theory to compute modal expansions of far-field quantities rigorously.
The presented approach is based on the
complex eigenfrequencies of QNMs.
The divergence problem is circumvented by
using contour integration with
an analytical continuation of the far-field quantity into the complex frequency plane.
We demonstrate the approach by computing the angular resolved modal energy flux in the far field of a nanophotonic device.
\end{abstract}
\maketitle

\section{Introduction}
Modern nanotechnology allows for exploring new regimes in tailoring light-matter
interaction~\cite{Novotny_NatPhot_2011}.
Applications comprise the design of nanoantennas for
quantum information technology~\cite{Ding_single_photons2016},
tuning photochemistry applications with nanoresonators~\cite{Zhang_ChemRev_2018},
using plasmonic nanoparticles for biosensing~\cite{Anker_BioSens_NatMater_2008},
and miniaturization of optical components using dielectric metasurfaces~\cite{Yu_NatMater_2014}.
Most approaches are based on
resonance phenomena. Optical resonances are characterized
by their wavelength-dependent localized and radiated field energies.
They may appear as, e.g., plasmonic resonances in metals~\cite{Tame_QuantPlasmon_NatPhys_2013}
or resonances in dielectric materials, such as Mie resonances~\cite{Kuznetsov_DielectricNanostruc_2016} or
bound states in the continuum~\cite{Hsu_BICs_NatRevMat_2016}.
The theoretical description of the resonances is essential
for understanding the physical properties of the systems and for
designing and optimizing related devices.
A popular approach is the modeling with QNMs,
which are the eigensolutions of resonant systems~\cite{Lalanne_QNMReview_2018,Kristensen_QNM_2020}.
In typical nanophotonic setups, outgoing radiation conditions have to be 
fulfilled yielding complex eigenfrequencies and an exponential decay 
of the QNMs in time. 
This means that the QNMs diverge exponentially with an
increasing distance to the 
resonators~\cite{Lamb_1900,Beck_1960,Lalanne_QNMReview_2018,Kristensen_QNM_2020}.
Due to the conceptual difficulties of exponential growth, this behavior has been termed ``exponential catastrophe''~\cite{Beck_1960}.
Nevertheless, QNM-based expansion approaches, where electromagnetic fields are expanded into
weighted sums of QNMs, have been derived to describe light-matter interaction in various
applications~\cite{Ching_RevModPhys_1998,Muljarov_EPL_2010,Vial_PRA_2014,Zolla_OptLett_2018,Zschiedrich_PRA_2018,Yan_PRB_2018}.
These approaches are based on the expansion of electromagnetic fields
inside and in the close vicinity of the resonators.
In this way, modal near-field quantities,
such as the modal Purcell
enhancement~\cite{Sauvan_QNMexpansionPurcell_2013,Zambrana-Puyalto_PRB_2015,Muljarov_ModalPurc_2016},
can be computed. For time-dependent problems, 
methods have been proposed to overcome
the divergence problem~\cite{Colom_PRB_2018,Abdelrahman_OSAConti_2018,Dezfouli_PRB_2018}.
\footnote{This work has been published:\\
F. Binkowski et al., Phys. Rev. B \textbf{102}, 035432 (2020).\\
DOI: \href{https://doi.org/10.1103/PhysRevB.102.035432}{10.1103/PhysRevB.102.035432}}

In many applications, time-averaged far-field quantities are of special
interest~\cite{Novotny_NatPhot_2011,Yu_NatMater_2014,Ding_single_photons2016}.
However, the divergence of QNMs is a key issue for modal
expansion of such quantities~\cite{Lalanne_QNMReview_2018,Kristensen_QNM_2020}.
From a physics perspective, 
for time-harmonic sources, the excited electromagnetic 
near- and far-field distributions are clearly nondiverging. 
This has motivated a discussion about the general 
applicability of QNMs~\cite{Chen_2019}.
Alternative approaches based on model approximations which yield eigenmodes
with real-valued frequencies in the far-field regions
have been proposed~\cite{Bergman_1979,Tureci_ConstantFlux_PRA_2006,Chen_2019}.
Further methods use the Dyson equation approach~\cite{Ge_NJOP_2014,Franke_PRL_2019} or near-field to far-field
transformations~\cite{Yang_ACS_Phot_2016} of the QNMs resulting in approximations
of the computed far-field quantities~\cite{Yan_PRB_2018,Ren_PRB_2020}.
Also, the intensively discussed question of how to normalize QNMs
is related to their exponential divergence
\cite{Kristensen_NormQNM_2015,Muljarov_Comment_PRA_2017,Kristensen_Reply_PRA_2017,Lalanne_QNMReview_2018,Stout_arXiv_2019,Kristensen_QNM_2020}.

In this work, 
we present a general approach for modal analysis which allows for
expansions of physical observables in the far-field region.
The approach
is based on the complex eigenfrequencies of the resonant systems; 
however, the diverging behavior of the corresponding QNMs is circumvented
by using contour integration of the relevant far-field quantities.
Therefore, the presented approach paves the way for avoiding 
an exponential catastrophe while retaining the rigorous model.
No approximation regarding the modeling of the
naturally complex-valued frequencies of a resonant system is required.
The method is validated by comparing the modal
expansion to a direct solution of the 
corresponding scattering problem. 
The approach is applied to compute
the modal expansion of the angular resolved energy flux density
radiated to the far field by 
a localized source in a resonant nanostructure.

\section{Modal expansion of far-field quantities}
\begin{figure}[]
	{\includegraphics[width=0.45\textwidth]{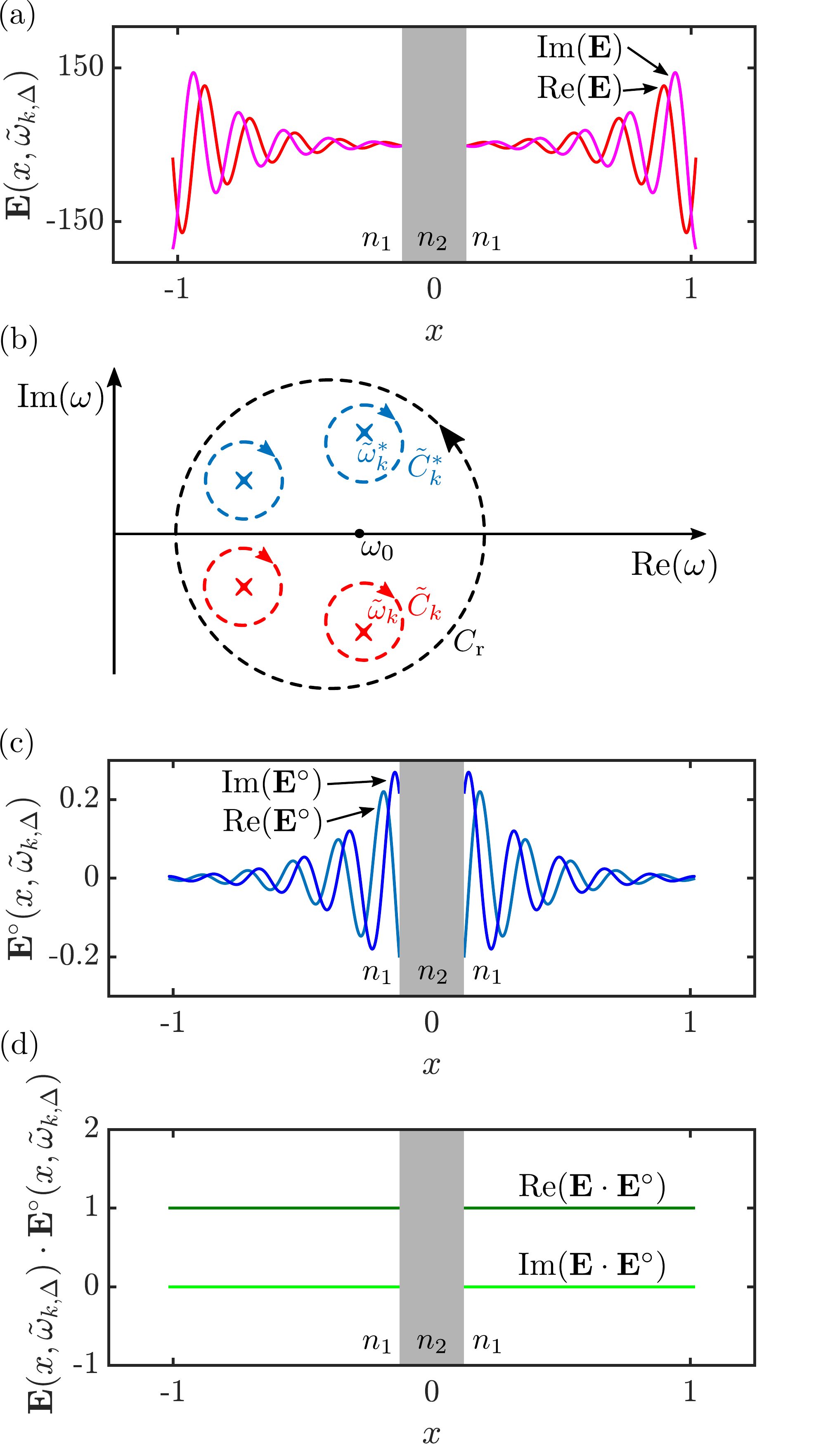}}
	\caption{One-dimensional resonator defined by layers with different refractive indices,
	where $n_2>n_1$. Electric field solutions, ${\mathbf{E}}(x,{\omega})$ and ${\mathbf{E}}^\circ(x,{\omega})$,
	are obtained by solving the Helmholtz equation with a source term corresponding to
	incoming plane waves with unit amplitude. Only scattered fields (a.u.) outside the resonator are shown.
	(a)~Diverging field
	${\mathbf{E}}(x,\tilde{\omega}_{k,\Delta}) =A e^{i (n_1\tilde{\omega}_{k,\Delta}/c) |x|}$,
	where $\tilde{\omega}_{k}$ is a resonance pole 
	of ${\mathbf{E}}(x,{\omega})$ and $\tilde{\omega}_{k,\Delta} = \tilde{\omega}_{k} + \Delta\tilde{\omega}_{k}$
	is a frequency close to $\tilde{\omega}_{k}$.
	(b)~Illustration of resonance poles and integration contours corresponding to
	 the RPE for the energy flux density given by~Eq.~\eqref{eq:full_RPE}. The analytical continuation
	 of the energy flux density has resonance poles with negative and with positive imaginary parts.
 	(c)~Nondiverging field
 	${\mathbf{E}}^\circ(x,\tilde{\omega}_{k,\Delta}) =B e^{-i (n_1\tilde{\omega}_{k,\Delta}/c) |x|}$.
 	(d)~Constant product
 	${\mathbf{E}}(x,\tilde{\omega}_{k,\Delta})\cdot{\mathbf{E}}^\circ(x,\tilde{\omega}_{k,\Delta})$,
 	which relates to the energy flux density.}
 \label{fig:fig01}
\end{figure} 
The QNMs of a resonant system are diverging outgoing waves.
Figure~\ref{fig:fig01}\textcolor{blue}{(a)} illustrates the electric field corresponding to a QNM in a
one-dimensional resonator defined by layers with different refractive indices.
In nano-optics, in the steady-state regime, electric fields $\mathbf{E}(\omega_0) \in \mathbb{C}^3$ are solutions to
the time-harmonic Maxwell's equations in second-order form,
\begin{align}
	\nabla \times \mu(\omega_0)^{-1} \nabla \times \mathbf{E}(\omega_0) -
	\omega_0^2\epsilon(\omega_0) \mathbf{E}(\omega_0) =
	i\omega_0\mathbf{J}, \label{eq:Maxwells_eq}
\end{align}
where $\omega_0\in\mathbb{R}$ is
the angular frequency and $\mathbf{J}\in \mathbb{C}^3$ is the source field.
For a simpler notation,
we omit the spatial dependence of the quantities and write, e.g.,
$\mathbf{E}(\omega_0)$ instead of
$\mathbf{E}(\mathbf{r},\omega_0)$, where
$\mathbf{r} \in \mathbb{R}^3$ is the position.
The permittivity tensor and the permeability tensor are defined by
$\epsilon(\omega_0)$ and $\mu(\omega_0)$, respectively.
For optical frequencies, $\mu(\omega_0)$ is
typically equal to the vacuum permeability $\mu_0$.
QNMs are solutions to Eq.~\eqref{eq:Maxwells_eq} equipped with outgoing
radiation conditions and without a source, i.e., $\mathbf{J} = 0$.
The eigenfrequencies $\tilde{\omega}_k \in \mathbb{C}$ have negative
imaginary parts and are given by the complex
resonance poles of the analytical continuation $\mathbf{E}(\omega)$ of
the electric field $\mathbf{E}(\omega_0)$ into the complex plane $\omega \in \mathbb{C}$.

We use the Riesz projection expansion (RPE)~\cite{Zschiedrich_PRA_2018,Binkowski_PRB_2019}
for modal expansion of the energy flux density in the far field,
which can be expressed as a quadratic form with a sesquilinear map.
The energy flux density~\cite{Jackson_3rdEd_1999} is defined by
\begin{align}
s(\mathbf{E}(\omega_0),\mathbf{E}^*(\omega_0)) \hspace{-0.025cm}=\hspace{-0.025cm}
\frac{1}{2} \mathrm{Re}\hspace{-0.05cm}\left(\hspace{-0.05cm}\mathbf{E}^*(\omega_0)
\hspace{-0.025cm}\times \hspace{-0.025cm}
\frac{1}{i \omega_0 \mu_0} \nabla \hspace{-0.025cm}\times \hspace{-0.025cm}
\mathbf{E}(\omega_0)\hspace{-0.075cm}\right)\hspace{-0.075cm}\cdot\hspace{-0.05cm}
\mathbf{n},\nonumber
\end{align}
where $\mathbf{E}^*(\omega_0)$ is the complex conjugate of the electric field and
$\mathbf{n}$ is the normal on the corresponding far-field sphere.
The RPE is based on contour integration in the complex frequency plane.
Since the complex conjugation of the electric field makes $s(\mathbf{E}(\omega_0),\mathbf{E}^*(\omega_0))$
nonholomorphic, the evaluation of this function for complex frequencies is problematic.
This challenge can be addressed by exploiting the relation
$\mathbf{E}^*(\omega_0) = \mathbf{E}(-\omega_0)$ for $\omega_0 \in \mathbb{R}$. The field $\mathbf{E}(-\omega_0)$
is a solution to Eq.~\eqref{eq:Maxwells_eq} as well.
For the harmonic time dependency $e^{-i\omega_0 t}$  with a negative frequency,
the radiation conditions are sign inverted.
The field $\mathbf{E}(-\omega_0)$ has an analytical continuation into the
complex plane $\omega\in \mathbb{C}$, which we denote by $\mathbf{E}^\circ(\omega)$. 
This yields the
required analytical continuation of
$s(\mathbf{E}(\omega_0),\mathbf{E}^*(\omega_0))$,
which is given by
$s(\mathbf{E}(\omega),\mathbf{E}^\circ(\omega))$. 
Note that $\mathbf{E}^\circ(\omega)$ introduces resonance poles in the upper
complex half-plane, which are usually not considered in the literature. These poles are an essential part of the presented approach.
To expand $s(\mathbf{E}(\omega_0),\mathbf{E}^*(\omega_0)) = s(\mathbf{E}(\omega_0),\mathbf{E}^\circ(\omega_0))$
into modal contributions, Cauchy's integral formula,
\begin{align}
	s(\mathbf{E}(\omega_0),\mathbf{E}^\circ(\omega_0)) =
	\frac{1}{2 \pi i} \oint \limits_{C_0} 
	\frac{s(\mathbf{E}(\omega),\mathbf{E}^\circ(\omega))}{\omega-\omega_0} \text{ d}\omega,
	\nonumber
\end{align}
is then exploited.
The contour
$C_0$ is a closed integration path around $\omega_0$ so that
$s(\mathbf{E}(\omega),\mathbf{E}^\circ(\omega))$ is holomorphic
inside of $C_0$. Deforming the integration path and applying Cauchy's residue theorem yield
\begin{align}
		\hspace{-0.12cm} s(\mathbf{E}(\omega_0),\mathbf{E}^\circ(\omega_0)) =
	  &- \sum_{k=1}^K \frac{1}{2 \pi i} \oint\limits_{{\tilde{C}}_k}
	\frac{s(\mathbf{E}(\omega),\mathbf{E}^\circ(\omega))}{\omega-\omega_0} \text{ d}\omega \nonumber \\
	& - \sum_{k=1}^K \frac{1}{2 \pi i} \oint\limits_{{\tilde{C}^*}_k}
	\frac{s(\mathbf{E}(\omega),\mathbf{E}^\circ(\omega))}{\omega-\omega_0} \text{ d}\omega \nonumber  \\
	&+ \frac{1}{2 \pi i}	\oint\limits_{C_{\text{r}}}
	\frac{	s(\mathbf{E}(\omega),\mathbf{E}^\circ(\omega))}{\omega-\omega_0} \text{ d}\omega, 
	\label{eq:full_RPE}
\end{align}
where $\tilde{C}_1,\dots,\tilde{C}_K$ are contours around 
the resonance poles of $\mathbf{E}(\omega)$, given by $\tilde{\omega}_1,\dots,\tilde{\omega}_K$,
and $\tilde{C}^*_1,\dots,\tilde{C}^*_K$ are contours
around the resonance poles
of $\mathbf{E}^\circ(\omega)$, given by
 $\tilde{\omega}^*_1,\dots,\tilde{\omega}^*_K$.
The outer contour $C_{\text{r}}$ includes $\omega_0$, the resonance poles
$\tilde{\omega}_1,\dots,\tilde{\omega}_K$ and $\tilde{\omega}^*_1,\dots,\tilde{\omega}^*_K$,
and no further poles, as sketched in 
Fig.~\ref{fig:fig01}\textcolor{blue}{(b)}.
The Riesz projections
\begin{align}
	{\tilde{s}}_k (\mathbf{E}(\omega_0),\mathbf{E}^\circ(\omega_0)) = &-\frac{1}{2 \pi i}
	\oint\limits_{{\tilde{C}}_k} \frac{s(\mathbf{E}(\omega),\mathbf{E}^\circ(\omega))}{\omega-\omega_0}
	\text{ d}\omega \nonumber \\
	&-\frac{1}{2 \pi i}
	\oint\limits_{{\tilde{C}^*}_k} \frac{s(\mathbf{E}(\omega),\mathbf{E}^\circ(\omega))}{\omega-\omega_0}
	\text{ d}\omega  \nonumber
\end{align}
are  modal contributions for the energy flux density.
The Riesz projections $\tilde{s}_k(\mathbf{E}(\omega_0),\mathbf{E}^\circ(\omega_0))$
are associated with the eigenfrequencies $\tilde{\omega}_k$
as the integration is performed along the contours $\tilde{C}_k$ and $\tilde{C}^*_k$.
The contribution
\begin{align}
	{{s}}_{\text{r}} (\mathbf{E}(\omega_0),\mathbf{E}^\circ(\omega_0)) = \frac{1}{2 \pi i}
	\oint\limits_{C_{\text{r}}}
	\frac{s(\mathbf{E}(\omega),\mathbf{E}^\circ(\omega))}{\omega-\omega_0} \text{ d}\omega \nonumber
\end{align}
is the remainder of the expansion containing nonresonant components as well as components corresponding to
eigenfrequencies outside of the contour $C_{\text{r}}$.
  
\begin{table}[]
	\begin{tabularx}{0.37\textwidth}{ccc} \mytoprule
		\hspace{0.2cm}$k$\hspace{0.2cm}
		&\hspace{0.1cm} $\mathrm{Re}(\tilde{\omega}_k)~[10^{15}\,\mathrm{s}^{-1}]$ \hspace{0.1cm}
		&\hspace{0.1cm} $\mathrm{Im}(\tilde{\omega}_k)~[10^{13}\,\mathrm{s}^{-1}]$ \hspace{0.1cm}\\
		\mymidrule
		$1$ & $1.441$ & - $0.109$ \\
		$2$ & $1.428$ & - $0.182$ \\
		$3$ & $1.399$ & - $0.232$ \\
		$4$ & $1.372$ & - $0.568$ \\
		$5$ & $1.370$ & - $1.025$ \\
		$6$ & $1.398$ & - $2.475$ \\
		$7$ & $1.406$ & - $0.470$ \\
		$8$ & $1.422$ & - $0.875$ \\
		$9$ & $1.435$ & - $1.942$ \\
		\mybottomrule
	\end{tabularx}
	\caption{Eigenfrequencies of the resonator shown in ~Fig.~\ref{fig:fig02}\textcolor{blue}{(a)}. 
	        The eigenfrequencies $\tilde{\omega}_k$ are contained in the circular contour $C_\mathrm{r}$,
		    which is centered at $1.41 \times 10^{15}\,\mathrm{s}^{-1}$ and has 
	        a radius of $6.8\times 10^{13}\,\mathrm{s}^{-1}$.}
	\label{tab:table1}
\end{table}  

\begin{figure}[h!]
	{\includegraphics[width=0.45\textwidth]{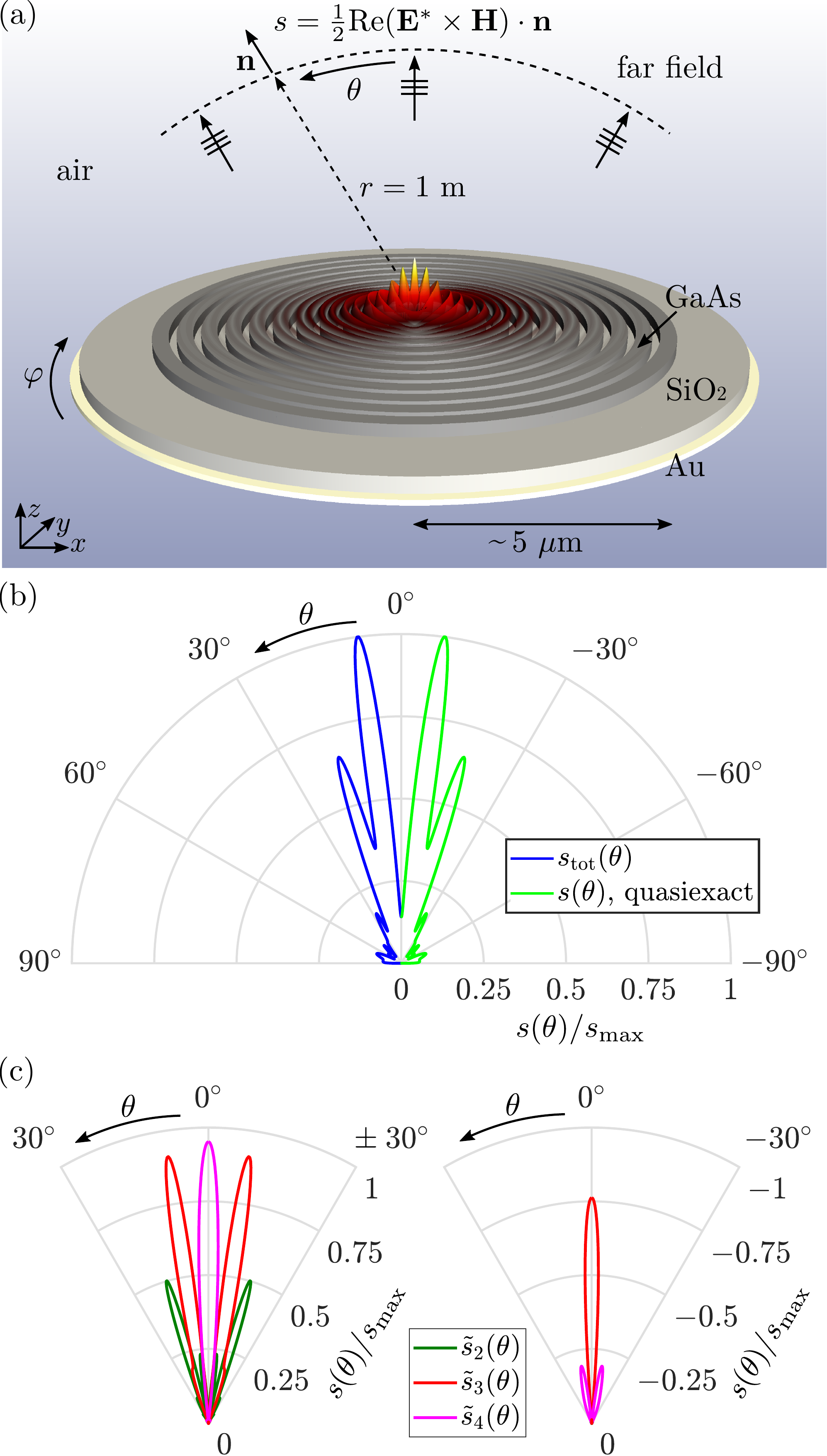}}
	\caption{Circular Bragg grating resonator with localized light source.
		(a)~Geometry with an illustration of the electric field intensity (a.u.) of the QNM
		corresponding to the eigenfrequency $\tilde{\omega}_2$; see Tab.~\ref{tab:table1}.
		The gallium arsenide (GaAs) grating has a thickness of
		$240\,\mathrm{nm}$ and consists of an inner disk with a radius of $550\,\mathrm{nm}$
		and 10 rings with a width of $340\,\mathrm{nm}$ and a periodicity of $500\,\mathrm{nm}$.
        The grating is placed on a silicon dioxide ($\mathrm{SiO}_2$) layer with a thickness of
        $240\,\mathrm{nm}$, which is coated from below with a gold (Au) layer of 
        $300\,\mathrm{nm}$ thickness.
		The light source is modeled by a dipole emitter placed at the center of the inner disk.
		The dipole radiates at the frequency $\omega_0$ and is oriented in $x$ direction.
		(b)~Radiation diagram at $\omega_0=2 \pi c/(1360\,\mathrm{nm})$ for the total modal expansion $s_\mathrm{tot}(\theta)$ 
		computed by~Eq.~\eqref{eq:full_RPE} and for the quasiexact solution of the energy flux density $s(\theta)$.
		The quantities are evaluated at
		$r = 1\,\mathrm{m}$ and $\varphi = 90^\circ$, which corresponds to the $yz$ plane.
		(c)~Modal decomposition of the radiation diagram for the contributions $\tilde{s}_2(\theta)$, $\tilde{s}_3(\theta)$, and $\tilde{s}_4(\theta)$.\vspace{-0.5cm}}
	\label{fig:fig02}
\end{figure} 
  
The RPE is based on evaluating ${{s}}(\mathbf{E}(\omega),\mathbf{E}^\circ(\omega))$
by solving Eq.~\eqref{eq:Maxwells_eq} for the frequencies $\omega$ and $-\omega$.
Consequently, the quadratic form
${{s}}(\mathbf{E}(\omega),\mathbf{E}^\circ(\omega))$,
where a product of $\mathbf{E}(\omega)$ and $\mathbf{E}^\circ(\omega)$ is involved,
does not diverge.
This is due to the cancellation of the factors $e^{i (n\omega/c) r}$ and $e^{-i (n\omega/c) r}$
of the fields in the far-field region, where $r = ||\mathbf{r}||$.
In this way, it becomes possible to compute modal expansions of
far-field quantities with nondiverging expansion terms.
To illustrate this, we consider a one-dimensional resonator and
compute electric fields, ${\mathbf{E}}(x,{\omega})$ and ${\mathbf{E}}^\circ(x,{\omega})$, fulfilling the
corresponding Helmholtz equation.
Figures~\ref{fig:fig01}\textcolor{blue}{(a)} and~\ref{fig:fig01}\textcolor{blue}{(c)} sketch
the diverging field ${\mathbf{E}}(x,\tilde{\omega}_{k,\Delta})$
and the nondiverging field ${\mathbf{E}}^\circ(x,\tilde{\omega}_{k,\Delta})$
outside of the resonator, respectively.
The frequency $\tilde{\omega}_{k,\Delta} = \tilde{\omega}_{k} + \Delta\tilde{\omega}_{k}$ represents an evaluation point
on an integration contour $\tilde{C}_k$. Figure~\ref{fig:fig01}\textcolor{blue}{(d)} shows
the nondiverging product
${\mathbf{E}}(x,\tilde{\omega}_{k,\Delta})\cdot{\mathbf{E}}^\circ(x,\tilde{\omega}_{k,\Delta})$,
which relates to the energy flux density.
The approach also applies to arbitrary three-dimensional problems, where,
in the far-field region,
${\mathbf{E}}(\mathbf{r},{\omega}) \sim e^{i (n\omega/c) r} (1/r)$ and 
${\mathbf{E}^\circ}(\mathbf{r},{\omega}) \sim e^{-i (n\omega/c) r} (1/r)$.

\section{Application}
The presented approach is used for modal analysis of a quantum technology device. 
We revisit an example from the literature~\cite{Rickert_OptExp_2019}, 
where a quantum dot acts as single-photon source. 
For a specific far-field region, the photon collection efficiency (PCE)
has been enhanced by using a numerically optimized circular Bragg grating nanoresonator. 
Such devices can be realized experimentally by using deterministic fabrication 
technologies~\cite{Senellart_2017}.
For more details on the specific device and material properties,
the reader is referred to~\cite{Rickert_OptExp_2019}.
The geometry is sketched in~Fig.~\ref{fig:fig02}\textcolor{blue}{(a)}.
To numerically analyze the light source,
we spatially discretize the system with the finite element method
(FEM) using the solver \textsc{JCMsuite}~\cite{Pomplum_NanoopticFEM_2007}. 

The quantity of interest is the energy flux density in the far field
$s(\omega_0,\theta) = s(\mathbf{E}(\omega_0,\theta),\mathbf{E}^\circ(\omega_0,\theta))$, see Eq.~\eqref{eq:full_RPE},
where $\theta$ is the inclination angle as shown in~Fig.~\ref{fig:fig02}\textcolor{blue}{(a)}.
For the modal expansion of $s(\omega_0,\theta)$,
the outer contour $C_\mathrm{r}$ is chosen to enclose the wavelength
range of interest, $1280\,\mathrm{nm} \leq \lambda_0 \leq 1400\,\mathrm{nm}$,
where $\lambda_0 = 2 \pi c/\omega_0$.
We compute all
eigenfrequencies inside of the contour, which are listed in Tab.~\ref{tab:table1}.
Note that only those rotationally symmetric QNMs which can couple to 
the dipole source are computed.
Figure~\ref{fig:fig02}\textcolor{blue}{(a)} sketches 
the electric field intensity of the QNM corresponding to $\tilde{\omega}_2$
in the near field of the structure.
The QNM exhibits a maximum of the field intensity at the center of the resonator 
and it diverges in the far-field region.

For a fixed dipole frequency, the radiation diagrams
for the total modal expansion
$s_\mathrm{tot}(\omega_0,\theta) =
\sum_{k=1}^9 \tilde{s}_k(\omega_0,\theta) + {s}_\mathrm{r}(\omega_0,\theta)$
and for the quasiexact solution $s(\omega_0,\theta)$
are depicted in Fig.~\ref{fig:fig02}\textcolor{blue}{(b)}.
The quasiexact solution is computed by solving scattering problems given by~Eq.~\eqref{eq:Maxwells_eq} directly.
The total modal expansion coincides with
the quasiexact solution with an absolute error of $s(\theta)/s_\mathrm{max}<5\times10^{-3}$
and, for the angle region $-60^\circ<\theta<60^\circ$, with a relative error smaller than $3\times 10^{-5}$.
The differences in these solutions are related to numerical discretization errors and would decrease further 
by refining the numerical parameters. 
The agreement demonstrates that, although the associated QNMs diverge in the far field,
the RPE of the energy flux density gives correct results
with nondiverging expansion terms. 
Figure~~\ref{fig:fig02}\textcolor{blue}{(c)}
shows the modal energy flux densities $\tilde{s}_2(\omega_0,\theta)$, $\tilde{s}_3(\omega_0,\theta)$, and
$\tilde{s}_4(\omega_0,\theta)$. 
These are the significant contributions 
for the total energy flux density and they have different directivities corresponding to the different 
diffraction intensities of the Bragg grating.
The contributions $\tilde{s}_3(\omega_0,\theta)$ and
$\tilde{s}_4(\omega_0,\theta)$ also have negative values.
A negative modal energy flux density can be understood as suppression of light emission into specific directions arising from 
the interference of various modes excited by the source at the frequency $\omega_0$.
Negative modal contributions have been reported also for QNM expansions 
of near-field quantities~\cite{Sauvan_QNMexpansionPurcell_2013}.
Note that, as physically expected, the total
modal expansion of the energy flux density,
$s_\mathrm{tot}(\omega_0,\theta)$, is positive for all angles~$\theta$.

\begin{figure}[]
	{\includegraphics[width=0.45\textwidth]{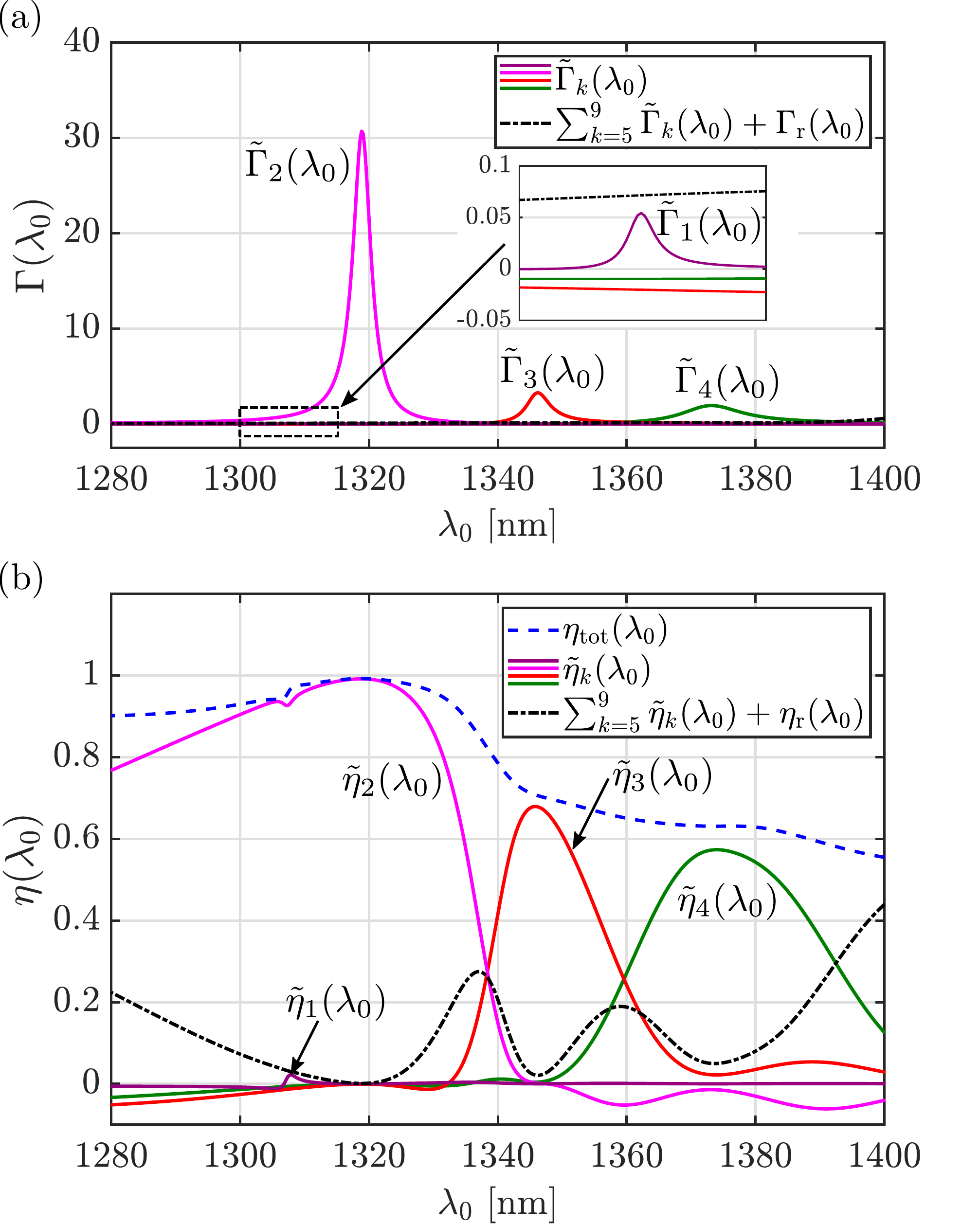}}
	\caption{Modal expansions of Purcell enhancement and PCE
	    for the resonator 
	    with a localized light source 
	    shown in ~Fig.~\ref{fig:fig02}\textcolor{blue}{(a)}.
		Eigenfrequencies $\tilde{\omega}_1,\dots,\tilde{\omega}_9$ are considered; see~Tab~\ref{tab:table1}.
		(a)~Modal expansion of the Purcell enhancement.
		The contributions $\tilde{\Gamma}_1(\lambda_0),\dots, \tilde{\Gamma}_4(\lambda_0)$ correspond to
		the eigenfrequencies $\tilde{\omega}_1,\dots,\tilde{\omega}_4$, respectively.
		The remaining modal contributions are added to the remainder of the expansion,
		$\sum_{k=5}^9 \tilde{\Gamma}_k(\lambda_0) + \Gamma_\mathrm{r}(\lambda_0)$. The term
		$\Gamma_\mathrm{r}(\lambda_0)$ includes also modal contributions
		corresponding to eigenfrequencies outside the integration contour $C_\mathrm{r}$.
		(b)~Modal expansion of the PCE.
		Total modal expansion, $\eta_\mathrm{tot}(\lambda_0) =
		\sum_{k=1}^9 \tilde{\eta}_k(\lambda_0) + {\eta}_\mathrm{r}(\lambda_0)$,
		single modal contributions, $\tilde{\eta}_1(\lambda_0),\dots, \tilde{\eta}_4(\lambda_0)$,
		and the sum of other contributions, $\sum_{k=5}^9 \tilde{\eta}_k(\lambda_0) + \eta_\mathrm{r}(\lambda_0)$.}
	\label{fig:fig03}
\end{figure} 

Next, the RPE is used to obtain 
insight into the properties of the device for the 
wavelength range $1280\,\mathrm{nm} \leq \lambda_0 \leq 1400\,\mathrm{nm}$.
Figure~\ref{fig:fig03}\textcolor{blue}{(a)} shows the
normalized decay rate, also termed Purcell enhancement, 
\begin{align}
	&\Gamma(\omega_0)= -\frac{1}{2} \mathrm{Re}(\mathbf{E}(\omega_0)\cdot \mathbf{j}^*)/\Gamma_\mathrm{b}, \nonumber
\end{align}
where $\mathbf{j}=-i\omega \mathbf{p}$ with the dipole moment $\mathbf{p}$
and $\Gamma_\mathrm{b}$ is the dipole emission in homogeneous background material~\cite{Zschiedrich_PRA_2018}.
It can be observed that, in the wavelength range of interest, the three resonances corresponding to the eigenfrequencies
$\tilde{\omega}_2$, $\tilde{\omega}_3$, and  $\tilde{\omega}_4$ are significant for the Purcell enhancement. 
The resonance with the eigenfrequency $\tilde{\omega}_1$ has a very small influence.
The nonresonant contributions and the contributions associated with other eigenfrequencies are negligible.
Figure~\ref{fig:fig03}\textcolor{blue}{(b)} shows the PCE,
\begin{align}
\eta(\omega_0)= \frac{1}{P_{\mathrm{DE}}}  \int_{\delta \Omega}
\frac{1}{2} \mathrm{Re}\left(\mathbf{E}^*(\omega_0) \times 
\frac{1}{i \omega_0 \mu_0} \nabla \times  \mathbf{E}(\omega_0)  \right)\cdot
\mathbf{\mathrm{d}S}, \nonumber
\end{align}
where $\delta \Omega$ is the far-field region corresponding
to $\mathrm{NA} = 0.8$ and $P_\mathrm{DE}$ is the emitted power of
the dipole emitter into the upper hemisphere. 
In the case of the PCE,
the resonances corresponding to
$\tilde{\omega}_1$,
$\tilde{\omega}_2$,
$\tilde{\omega}_3$, and
$\tilde{\omega}_4$
play an important role.
In contrast to the Purcell enhancement, the modal contribution $\tilde{\eta}_1(\omega_0)$
is significant for the PCE.
It contributes to $\eta_{\mathrm{tot}}(\omega_0)$
for the wavelength region near to its maximum.
Note that the behavior of the remaining contributions,
$\sum_{k=5}^9 \tilde{\eta}_k(\lambda_0) + \eta_\mathrm{r}(\lambda_0)$,
is partially based on resonances with eigenfrequencies outside the
integration contour $C_\mathrm{r}$. 

\section{Conclusions}
A theoretical approach to investigate
modal quantities in the far field of resonant systems was presented. 
Although the QNMs decay exponentially in time and thus represent diverging outgoing waves,
modal expansions can be computed rigorously.
The approach was applied to expand
the energy flux density in the far field of a nanoresonator with an embedded point source.
It was demonstrated that, by computing modal far-field patterns, 
those resonances which contribute significantly to the scattering response 
of the nanophotonic device can be identified. 
Thus deeper physical insights into the system are gained. 

The method cannot only be used to efficiently compute the 
scattering response and to compare to experimental results, but also 
for an optimization of devices for a tailored functionality. 
It can be applied to far-field as well as to near-field quantities.
Examples are quantities involving the electromagnetic energy flux density or the electromagnetic absorption.
However, the investigations in this work are limited to quadratic forms with a sesquilinear map.
We expect
that, with resolving the key issue of the far-field treatment
in QNM modeling, the presented approach will enable usage
of QNMs in various fields. Applications include systems in
nano-optics with any material dispersion and any resonant
system in general, e.g., in acoustics or quantum mechanics.

\section*{Acknowledgments}
We acknowledge funding by the Deutsche Forschungsgemeinschaft
(DFG, German Research Foundation) under Germany’s Excellence Strategy -- The Berlin Mathematics
Research Center MATH+ (EXC-2046/1, Project ID No. 390685689, AA4-6).
We acknowledge the Helmholtz Association for funding within the Helmholtz Excellence
Network SOLARMATH, a strategic collaboration of the DFG
Excellence Cluster MATH+ and Helmholtz-Zentrum Berlin
(Grant No. ExNet-0042-Phase-2-3). This work is partially
funded through the project 17FUN01 “BeCOMe” within the
Programme EMPIR. The EMPIR initiative is co-founded by
the European Union’s Horizon 2020 research and innovation
program and the EMPIR Participating Countries.

\end{document}